\documentclass[showpacs]{revtex4}
\usepackage{amsmath}
\usepackage{amssymb}
\usepackage{graphicx}

\begin{document}
\title{
Can mixed star-plus-wormhole systems mimic black holes?
 }
\author{
Vladimir Dzhunushaliev,$^{1,2,3,4}$
\footnote{v.dzhunushaliev@gmail.com}
Vladimir Folomeev,$^{1,4}$
\footnote{vfolomeev@mail.ru}
Burkhard Kleihaus,$^{5}$
\footnote{b.kleihaus@uni-oldenburg.de }
Jutta Kunz$^5$
\footnote{jutta.kunz@uni-oldenburg.de}
}
\affiliation{
$^1$
Institute for Basic Research, Eurasian National University, Astana 010008, Kazakhstan\\
$^2$
Department of Theoretical and Nuclear Physics,  Al-Farabi Kazakh National University, Almaty 050040, Kazakhstan \\
$^3$ Institute of Experimental and Theoretical Physics,  Al-Farabi Kazakh National University, Almaty 050040, Kazakhstan \\
$^4$Institute of Physicotechnical Problems and Material Science of the NAS
of the Kyrgyz Republic, 265 a, Chui Street, Bishkek 720071,  Kyrgyz Republic \\
$^5$Institut f\"ur Physik, Universit\"at Oldenburg, Postfach 2503
D-26111 Oldenburg, Germany
}

\begin{abstract}
We consider mixed strongly gravitating configurations consisting of a wormhole
threaded by two types of ordinary matter. For such systems, the possibility of obtaining
static spherically symmetric solutions describing
compact massive central objects enclosed by high-redshift surfaces (black-hole-like configurations) is studied.
Using the standard thin accretion disk model, we exhibit potentially observable differences allowing to distinguish
the mixed systems from ordinary black holes with the same masses.
\end{abstract}

\maketitle
\section{Introduction}

It is now commonly believed that objects described by solutions with an event horizon -- black holes (BH) --
can exist in the Universe~\cite{Shapiro:2008}. In the simplest case it can be the
 well-known exact spherically
symmetric solution found by Schwarzschild in 1916, immediately after the creation of
Einstein's general relativity. However, even 100 years later,
the physical reality of such a mathematical solution is still sometimes questioned, essentially
because it is so far not possible to prove unambiguously that
astrophysical BH candidates really possess an  event horizon \cite{Abramowicz:2002vt,Narayan:2005ie}.
Therefore, it seems that there is no {\it a priori} reason to exclude from consideration
astrophysical objects that do not have an event horizon but are able to
mimic the main observational characteristics of BHs.

In this connection, the literature in the field offers alternative types of configurations
which, from the point of view of a distant observer, would look almost like BHs
but would have no horizon and singularity -- the so-called black hole mimickers (BHM).
Among them are boson stars~\cite{Torres:2002td,Guzman:2009zz},
gravastars~\cite{Mazur:2001fv,Visser:2003ge,Harko:2009gc}, and
wormholes~\cite{Damour:2007ap,Lemos:2008cv,Harko:2008vy}.
Obviously,
the properties of gravitational fields produced by such objects
will depend on the particular  BHM configuration. Then it will manifest itself, for example, in considering the process of accretion
of matter onto objects of this kind.  Because of the differences in their external geometry,
one might expect changes both in the structure of accretion disks
and in their emission spectra.

In the present paper we consider a mixed relativistic configuration consisting
of a wormhole filled by ordinary matter.
In this way  we suggest one more possible way to mimic a ``black-hole-like''
configuration. (By the latter, as in the case of other objects of this kind,
we mean a compact massive central object enclosed by a high-redshift surface.)
In our previous
works~\cite{Dzhunushaliev:2011xx,Dzhunushaliev:2012ke,Dzhunushaliev:2013lna,Charalampidis:2013ixa,Hauser:2013jea,Dzhunushaliev:2014bya,
Dzhunushaliev:2014mza,Aringazin:2014rva,Dzhunushaliev:2015sla}
we have already studied systems of this kind in various aspects of the problem.
In particular, using the obtained equilibrium neutron-star-plus-wormhole configurations,
some issues concerning possible astrophysical manifestations associated with the presence of nontrivial topology in the system
have been considered. Namely, in Ref.~\cite{Dzhunushaliev:2014mza}
the passage of light -- radiated from the surface of a neutron star --
through a throat of a wormhole has been studied.
It was shown that in this case there is  a
characteristic  distribution of the intensity of the light which differs
from the one obtained when
considering the case where radiation does not pass through the throat. In
principle, such an effect could be observed by instruments with sufficiently high resolution.
In Ref.~\cite{Aringazin:2014rva} the influence  of the nontrivial topology on
the structure of the interior magnetic field of mixed systems supported by neutron matter modeled
by isotropic and anisotropic fluids have been demonstrated.
Here we would like to continue searching for other potentially observable astrophysical manifestations shared by mixed star-plus-wormhole systems.
In the process, our purpose will be twofold: (i) we demonstrate the possibility that BHM solutions can be obtained in such mixed systems;
(ii) we reveal potentially observable effects which distinguish such systems from other
BHMs considered in the literature.

In doing so, we first construct static spherically symmetric solutions whose nontrivial topology is provided by the presence of a
ghost scalar field. This type of fields are now widely used in modeling  the accelerated expansion of the present Universe~\cite{AmenTsu2010}.
 With the opposite sign in front of its kinetic energy term, such a field violates
the null energy condition that may lead to the appearance of a nontrivial wormholelike topology.
The aim of the present work
is to study possible observational differences between
mixed systems and ordinary BHs with the same masses.

The important observational manifestations of BHs are the effects associated with a process of accretion of
surrounding matter onto a BH. For thin accretion
disks, the energy released in such a process
is $\sim 6\%$ to 42\% (depending on the spin of the black hole) of the rest mass of the accreting matter,
and this energy may be converted into observable radiation~\cite{Shapiro:2008}.
Calculations of the accretion flow onto a BH and the emitted radiation pattern are, in general, very difficult.
But since our purpose here is just to reveal the differences between the accretion onto BHs and our mixed systems but
not a more or less realistic modeling of the accretion process in itself, we restrict ourselves  to the consideration
of a relatively simple model. Namely,
 we will consider a steady-state accretion process for a geometrically thin and optically thick accretion disc
orbiting the mixed configurations. To reveal the differences,
we will compare our results with those obtained for BHs with the same masses.

The paper is organized as follows.
In Sec.~\ref{statem_prob} we present the general-relativistic  equations for the mixed systems under consideration
and describe two particular choices of the equation of state for ordinary matter.
In order to obtain black-hole-like solutions,
in Sec.~\ref{num_calc} we numerically solve these equations
with different choices for the parameters of the systems.
To demonstrate the observational differences between the obtained mixed  systems and ordinary BHs,
in Sec.~\ref{sec_thin_acc_disk} we consider  the process of thin-disk accretion onto such configurations
and compare the energy fluxes emitted  from the disk's surface.
Finally,  in Sec.~\ref{conclusion} we discuss and summarize the obtained results.

\section{Statement of the problem}
\label{statem_prob}

We will consider a mixed system containing two types of fluid: (i) an ordinary fluid satisfying all
energy conditions, and (ii) an exotic fluid violating the null energy condition.
For the ordinary fluid, one can take in principle any form of fluid.
For example, it could be ordinary matter that stars are made of,
including neutron stars~\cite{Dzhunushaliev:2011xx,Dzhunushaliev:2012ke,Dzhunushaliev:2013lna,Dzhunushaliev:2014mza,Aringazin:2014rva,Dzhunushaliev:2015sla},
or dark matter, but it could also comprise electromagnetic fields~\cite{Kardashev:2006nj,Novikov:2007zz},
chiral fields~\cite{Charalampidis:2013ixa}, Yang-Mills fields~\cite{Hauser:2013jea}, or complex scalar fields~\cite{Dzhunushaliev:2014bya}.

As regards the exotic fluid, we will consider a situation where its presence
gives rise to a nontrivial wormholelike spacetime topology of the system. Modeling of such a fluid can be done
in many ways,  both within the frameworks of general relativity and when considering modified theories
of gravity.

\subsection{General equations}

To demonstrate a possibility of obtaining  black-hole-like solutions for the aforementioned mixed systems,
let us consider a situation where:

(i) Ordinary matter is modeled by an isotropic perfect fluid, i.e., by a fluid with equal radial and tangential pressures.
Its energy-momentum tensor is
\begin{equation}
\label{emt_perf_fluid}
T_{i\text{(fl)}}^k=(\varepsilon+p)u_i u^k-\delta_i^k p~,
\end{equation}
where $\varepsilon$, $p$, and $u^i$ are the energy density, the pressure,
and the four-velocity of the fluid, respectively.

(ii) Exotic matter is described by one ghost scalar field $\varphi$,
i.e., by a field with the opposite sign in front of its kinetic energy term, with the following
energy-momentum tensor:
 \begin{equation}
\label{emt_sf}
T_{i\text{(sf)}}^k=-\partial_{i}\varphi\partial^{k}\varphi
-\delta_i^k\left[-\frac{1}{2}\partial_{\mu}\varphi\partial^{\mu}\varphi-V(\varphi)\right]~,
\end{equation}
where $V(\varphi)$ is the potential energy.

A necessary condition for providing a nontrivial wormhole topology in the system is the violation of the null energy condition, $T_{ik}n^i n^k \geq 0$,
where $T_{ik}=T_{i\text{(fl)}}^k+T_{i\text{(sf)}}^k$ and $n^i$ is any null vector.
In our case this implies that the following inequalities are satisfied (at least in some region of spacetime): $T_0^0-T_1^1<0, T_1^1>0$.

For the mixed system under consideration, it is convenient to use polar Gaussian coordinates. The metric then reads
 \begin{equation}
\label{metric_wh_poten}
ds^2=e^{\nu}(dx^0)^2-dr^2-R^2 d\Omega^2,
\end{equation}
where $\nu$  and $R$ are functions of the radial coordinate $r$ only,
 $d\Omega^2$ is the metric on the unit two-sphere, and the time coordinate  $x^0=c\, t$.
 Then the corresponding components of the energy-momentum tensor take the form
 \begin{eqnarray}
\label{emt-00}
&&T_0^0=\varepsilon-\frac{1}{2}\varphi^{\prime 2}+V,\\
\label{emt-11}
&&T_1^1=-p+\frac{1}{2}\varphi^{\prime 2}+V,\\
\label{emt-22}
&&T_2^2=T_3^3=-p-\frac{1}{2}\varphi^{\prime 2}+V.
 \end{eqnarray}

We will consider here the simplest situation
when ordinary matter is located  in the central region of the system.
The center is described by the value of the radial coordinate $r=0$,
which corresponds to a throat or an equator.
The term center thus refers to the extremal surface,
located symmetrically between the two asymptotically flat regions.
We also assume that the matter has
a maximum  density at the center.
(For cases of a shifted maximum density,
see Refs.~\cite{Aringazin:2014rva,Dzhunushaliev:2015sla}.) Also, without
loss of generality, we can set the value of the scalar field at the center to
$\varphi(0)=0$, but we note that $\varphi^{\prime}(0)\neq 0$.
Then the potential of the scalar field can be expanded  in the neighborhood of the center as
 \begin{equation}
				\label{expan_sf}
\varphi \approx \varphi_1 r +\frac{1}{6}\varphi_3 r^3,
\end{equation}
where $\varphi_1$ is the derivative at the center,
 the square of which corresponds to the ``kinetic'' energy of the scalar field.
In further calculations we will use this kinetic energy to introduce dimensionless variables.

Taking into account the components of the fluid energy-momentum tensor  \eqref{emt-00}-\eqref{emt-22},
the $(_0^0)$, $(_1^1)$, and $(_2^2)$ components of the Einstein equations with the metric
 \eqref{metric_wh_poten}  take the form
\begin{eqnarray}
\label{Einstein-00_poten}
&&-\left[2\frac{R^{\prime\prime}}{R}+\left(\frac{R^\prime}{R}\right)^2\right]+\frac{1}{R^2}
=\frac{8\pi G}{c^4} T_0^0,
 \\
\label{Einstein-11_poten}
&&-\frac{R^\prime}{R}\left(\frac{R^\prime}{R}+\nu^\prime\right)+\frac{1}{R^2}
=\frac{8\pi G}{c^4} T_1^1,
\\
\label{Einstein-22_poten}
&&\frac{R^{\prime\prime}}{R}+\frac{1}{2}\frac{R^\prime}{R}\nu^\prime+
\frac{1}{2}\nu^{\prime\prime}+\frac{1}{4}\nu^{\prime 2}
=-\frac{8\pi G}{c^4} T_2^2,
\end{eqnarray}
where the prime denotes  differentiation with respect to $r$.

The general equation for the scalar field $\varphi$  is
\begin{equation}
\label{sf_eq_gen}
\frac{1}{\sqrt{-g}}\frac{\partial}{\partial x^i}\left(\sqrt{-g}g^{ik}\frac{\partial \varphi}{\partial x^k}\right)=\frac{d V}{d \varphi}.
\end{equation}
Using the metric \eqref{metric_wh_poten}, this equation gives
\begin{equation}
\label{sf_poten}
\varphi^{\prime\prime}+\left(\frac{1}{2}\nu^\prime+2\frac{R^\prime}{R}\right)\varphi^\prime=-\frac{d V}{d \varphi}.
\end{equation}

Because of the conservation of energy and momentum,
$T^k_{i; k}=0$,
not all of the Einstein field equations are independent.
Taking the $i=1$ component of the conservation equations gives
\begin{equation}
\label{conserv_1}
\frac{d T^1_1}{d r}+
\frac{1}{2}\left(T_1^1-T_0^0\right)\nu^\prime+2\frac{R^\prime}{R}\left[T_1^1-\frac{1}{2}\left(T^2_2+T^3_3\right)\right]=0.
\end{equation}
Taking into account the components \eqref{emt-00}-\eqref{emt-22},
and also Eq.~\eqref{sf_poten}, we obtain from Eq.~\eqref{conserv_1}
\begin{equation}
\label{conserv_2}
\frac{d p}{d r}=-\frac{1}{2}(\varepsilon+p)\frac{d\nu}{d r}.
\end{equation}

Keeping in mind that the pressure and the energy density of ordinary matter are related by some equation of state (EOS),
 we have four unknown functions~-- $R$, $\nu$, $p$, and $\varphi$~-- for which there are five equations,
\eqref{Einstein-00_poten}-\eqref{Einstein-22_poten},
\eqref{sf_poten}, and \eqref{conserv_2}, only four of which are independent.
We will consider below two types of EOS, used both in describing compact astrophysical objects and
in modeling dark matter. In both cases, for simplicity, we assume that the scalar field is massless and has no self-interaction,
i.e., that $V=0$.

\subsection{Polytropic EOS}
\label{sec_poly_eos}
Consider first the case of ordinary (neutron) matter modeled by a polytropic EOS.
Such an EOS, being, on one hand, relatively simple, 
reflects adequately the  general properties of more realistic EOSs describing matter at small and
high densities and pressures.
This EOS can be taken in the following form~\cite{Tooper2}:
\begin{equation}
\label{eqs_NS_WH}
p=K \rho_{b}^{1+1/n}, \quad \varepsilon = \rho_b c^2 +n p,
\end{equation}
with the constant $K=k c^2 (n_{b}^{(ch)} m_b)^{1-\gamma}$,
and the polytropic index $n=1/(\gamma-1)$,
and where $\rho_b=n_{b} m_b$ denotes the rest-mass density
of the neutron fluid. Here $n_{b}$ is the baryon number density,
$n_{b}^{(ch)}$ is a characteristic value of $n_{b}$,
$m_b$ is the baryon mass,
and $k$ and $\gamma$ are parameters
whose values depend on the properties of the neutron matter.

The literature in the field offers a variety of values for the parameters entering this EOS.
This allows the possibility of getting both weakly and  strongly relativistic objects~\cite{Tooper2}.
For simplicity, here we take only one set of parameters for the neutron fluid.
Namely, we choose
$m_b=1.66 \times 10^{-24}\, \text{g}$,
$n_{b}^{(ch)} = 0.1\, \text{fm}^{-3}$,
$k=0.1$, and $\gamma=2$ \cite{Salg1994}.
We employ these values for the parameters in the numerical calculations of Sec.~\ref{num_calc}.

To carry out numerical calculations, it is convenient to
rewrite the above equations in terms of dimensionless variables.
This can be done as follows:
\begin{equation}
\label{dimless_xi_v}
\xi=\frac{r}{L}, \quad \Sigma=\frac{R}{L},
\quad \phi(\xi)=\frac{\sqrt{8\pi G}}{c^2}\,\varphi(r)
\quad \text{with} \quad L=\frac{c^2}{\sqrt{8\pi G}\varphi_1},
\end{equation}
where $L$ is the characteristic size of the system. In turn,
 for  the fluid density one can use the new reparametrization~\cite{Zeld},
\begin{equation}
\label{theta_def}
\rho_b=\rho_{b c} \theta^n~,
\end{equation}
where $\rho_{b c}$ is the density of the neutron fluid at the
center of the configuration. Then Eqs.~\eqref{Einstein-00_poten}-\eqref{Einstein-22_poten},
 \eqref{sf_poten}, and \eqref{conserv_2} take the following dimensionless form:
\begin{eqnarray}
\label{Einstein-00_dmls}
&&-\left[2\frac{\Sigma^{\prime\prime}}{\Sigma}+\left(\frac{\Sigma^\prime}{\Sigma}\right)^2\right]+\frac{1}{\Sigma^2}
=\tilde T_0^0,
 \\
\label{Einstein-11_dmls}
&&-\frac{\Sigma^\prime}{\Sigma}\left(\frac{\Sigma^\prime}{\Sigma}+\nu^\prime\right)+\frac{1}{\Sigma^2}
=\tilde T_1^1,
\\
\label{Einstein-22_dmls}
&&\frac{\Sigma^{\prime\prime}}{\Sigma}+\frac{1}{2}\frac{\Sigma^\prime}{\Sigma}\nu^\prime+
\frac{1}{2}\nu^{\prime\prime}+\frac{1}{4}\nu^{\prime 2}
=-\tilde T_2^2,
\\
\label{sf_dmls}
&&\phi^{\prime 2}=\frac{e^{\nu_c-\nu}}{(\Sigma/\Sigma_c)^4},
\\
\label{fluid_dmls}
&&\sigma(n+1)\theta^\prime+\frac{1}{2}\left[1+\sigma (n+1)\theta\right]\nu^\prime=0.
\end{eqnarray}
Here the dimensionless right-hand sides of the Einstein equations are:
\begin{equation}
\label{emt_comp_dmls_poly}
\tilde T_0^0=B  (1+\sigma n \theta) \theta^n
-\frac{1}{2}\phi^{\prime 2}, \quad
\tilde T_1^1=-B  \sigma  \theta^{n+1}
+\frac{1}{2}\phi^{\prime 2}, \quad
\tilde T_2^2=-B \sigma  \theta^{n+1}
-\frac{1}{2}\phi^{\prime 2},
\end{equation}
where
$B=(\rho_{b c} c^2)/\varphi_1^2$ is the dimensionless ratio of the fluid energy density to that
of the scalar field at the center; $\Sigma_c$ and $\nu_c$ are the values of the corresponding functions at the center
			 [see Eq.~\eqref{bound_stat}];
$\sigma=K \rho_{b c}^{1/n}/c^2=p_c/(\rho_{b c} c^2)$ is a constant,
related to the pressure $p_c$ of the fluid at the center. The values of the fluid parameters appearing here
are taken from the above text [see after Eq.~\eqref{eqs_NS_WH}].
Eq.~\eqref{sf_poten} has been integrated to give the expression~\eqref{sf_dmls} with the integration constant chosen so as to
provide $\phi^\prime=1$ at the center.

 Eq.~\eqref{fluid_dmls} may be integrated to
give in the internal region with $\theta \ne 0$ the metric function $e^{\nu}$
in terms of $\theta$,
\begin{equation}
\label{nu_app}
e^{\nu}=e^{\nu_c}\left[\frac{1+\sigma (n+1)}{1+\sigma (n+1)\theta}\right]^{2},
\end{equation}
and $e^{\nu_c}$ is the value of $e^{\nu}$ at the center where $\theta=1$.
		   The integration constant $\nu_c$ is fixed
by requiring that the space-time is asymptotically flat,
i.e., $e^{\nu}=1$ at infinity.

\subsubsection{Boundary conditions}

We here consider neutron-star-plus-wormhole configurations
that are asymptotically flat and
symmetric under $\xi \to -\xi$.
The metric function $\Sigma(\xi)$ may be considered
as a dimensionless circumferential radial coordinate.
Asymptotic flatness requires that $\Sigma(\xi) \to |\xi|$ for large $|\xi|$.
Because of the assumed symmetry of the configurations,
the center of the configurations at
		    $\xi=0$ should correspond to an extremum
of $\Sigma(\xi)$, i.e.,~$\Sigma'(0)=0$.
If $\Sigma(\xi)$ has a minimum at $\xi=0$, then $\xi=0$ corresponds to the throat
of the wormhole.
If, on the other hand, $\Sigma(\xi)$ has a local maximum at $\xi=0$,
then $\xi=0$ corresponds to an equator.
In that case, the wormhole will have a double throat
surrounding a belly (see, e.g., Refs.~\cite{Charalampidis:2013ixa,Hauser:2013jea,Dzhunushaliev:2014mza}).

Expanding the metric function
$\Sigma$ in the neighborhood of the center
$$\Sigma\approx \Sigma_c+1/2\, \Sigma_2 \xi^2$$
 and using Eqs.~\eqref{Einstein-00_dmls} and \eqref{Einstein-11_dmls}, we find
the relations
\begin{equation}
\label{bound_coef}
\Sigma_c=\frac{1}{\sqrt{1/2-B \sigma}}, \quad \Sigma_2=\frac{\Sigma_c}{2}\Big\{1-B\left[1+\sigma(n+1)\right]\Big\}.
\end{equation}
Thus the sign of the expansion coefficient $\Sigma_2$ determines whether
the configurations possess a single throat at the center or an equator
surrounded by a double throat.

Equations \eqref{Einstein-00_dmls}-\eqref{fluid_dmls} are solved
for given parameters of the fluid $\sigma$, $n$, and $B$,
subject to the boundary conditions at the center of the configuration $\xi=0$,
\begin{equation}
			       \label{bound_stat}
\Sigma(0)=\Sigma_c, \quad \Sigma^\prime (0)=0,
\quad \nu(0)=\nu_c, \quad
\phi(0) =0, \quad \phi^\prime (0)=1,
\end{equation}
and also $\theta(0) = 1$.
Note here that, using \eqref{dimless_xi_v}, we can express the dimensional value of the derivative $\varphi_1$  as follows:
\begin{equation}
\label{kinet_scal}
\varphi_1^2=\frac{c^4}{8\pi G}\frac{1}{L^2}.
\end{equation}
Thus the dimensional ``kinetic'' energy of the scalar field depends only on the value of the characteristic length $L$,
which can be chosen arbitrarily subject to some physically reasonable assumptions.
Substituting this $\varphi_1^2$ into
the expression for $B$ [see after Eq.~\eqref{emt_comp_dmls_poly}], we find
\begin{equation}
\label{B_poly}
B=8\pi G \rho_{b c} (L/c)^2.
\end{equation}
It is seen
from the above expressions for $\varphi_1^2$ and $B$ that by fixing $L$,
one automatically determines the value of $\varphi_1^2$.
But the value of $B$ can still change depending on
the value of the fluid density $\rho_{b c}$ at the center.
Therefore one can consider $B$ as a parameter
describing the ratio of the fluid energy density at the center
to the energy density of the scalar field at the center.

Aside from giving the boundary conditions at the center,  it is important for us to keep track of the behaviour of the system on the other boundary -- the surface of the fluid.
Like the characteristics of the central region, the properties of the fluid's boundary  (in particular, magnitudes of the surface red shift) will also be determined by the parameters of the system.
For more discussion of this issue, see Sec.~\ref{num_calc}.

\subsection{Completely degenerate Fermi gas}

One more simple type of EOS used in the literature to model compact objects
(white dwarfs and neutron stars~\cite{Shapiro:2008}, dark matter stars~\cite{Narain:2006kx}) is an EOS describing an ideal
 completely degenerate Fermi gas at zero temperature.
Its equation of state can be obtained by using usual expressions for the energy density and pressure~\cite{Shapiro:2008}:
\begin{eqnarray}
\label{eos_enrg_dens}
&&\varepsilon=\frac{1}{\pi^2 \hbar^3}\int_0^{k_F}k^2\sqrt{m_f^2 c^4+k^2 c^2}dk=\frac{c^5 m_f^4}{\hbar^3}\frac{1}{8\pi^2}\left[z\sqrt{1+z^2}(1+2 z^2)-\sinh^{-1}(z)\right]\equiv \frac{c^5 m_f^4}{\hbar^3} \chi_1
,\\
\label{eos_p}
&&p=\frac{1}{3\pi^2 \hbar^3}\int_0^{k_F}\frac{k^4 c^2}{\sqrt{m_f^2 c^4+k^2 c^2}}dk=\frac{c^5 m_f^4}{\hbar^3}\frac{1}{8\pi^2}\left[z\sqrt{1+z^2}(2/3 \,z^2-1)+ \sinh^{-1}(z)\right]\equiv \frac{c^5 m_f^4}{\hbar^3} \chi_2,
\end{eqnarray}
where $\chi_1, \chi_2$ are the dimensionless energy density and pressure, respectively.
Here $m_f$ is  the fermion mass, $k_F$ is  the Fermi momentum, $z=k_F/(m_f c)$  is the relativity parameter.
Eqs.~\eqref{eos_enrg_dens} and \eqref{eos_p} yield  a parametric dependence $p=p(\varepsilon)$.

In two limiting cases, this EOS can be represented in simple power-law forms:
(i) in the nonrelativistic case, $z\ll 1$, we get the polytropic law, $\chi_2 \propto \chi_1^{5/3}$, and
(ii) in the ultrarelativistic case, $z\gg 1$, we have $\chi_2 = \chi_1/3$.

Using this EOS and the dimensionless variables \eqref{dimless_xi_v}, we get
for the right-hand sides of Eqs.~\eqref{Einstein-00_dmls}-\eqref{Einstein-22_dmls}
\begin{equation}
\label{emt_comp_dmls_fermi}
\tilde T_0^0=B_1  \chi_1
-\frac{1}{2}\phi^{\prime 2}, \quad
\tilde T_1^1=-B_1  \chi_2
+\frac{1}{2}\phi^{\prime 2}, \quad
\tilde T_2^2=-B_1 \chi_2
-\frac{1}{2}\phi^{\prime 2},
\end{equation}
where
$B_1=\frac{c^5 m_f^4}{\hbar^3  \varphi_1^2}\equiv 8\pi G L^2 \frac{ m_f^4 c}{\hbar^3} $.
In turn, instead of Eq.~\eqref{fluid_dmls}, we have
\begin{equation}
\label{ferm_fluid_dmls}
\frac{d\chi_1}{d\xi}=-\frac{1}{2}\frac{\chi_1+\chi_2}{d\chi_2/d\chi_1}\frac{d\nu}{d\xi}.
\end{equation}

\subsubsection{Boundary conditions}

As before, we choose the boundary conditions in the form of \eqref{bound_stat}.
Then, taking into account the expansion in the vicinity of the center,
\begin{equation}
\label{bound_chi}
\chi_1\approx \chi_{1 c}+\frac{1}{2}\chi_{12}\xi^2, \quad
\chi_2\approx \chi_{2 c}+\frac{1}{2}\chi_{22}\xi^2,
\end{equation}
where $\chi_{1 c}, \chi_{2 c}$ are the values of the dimensionless energy density and pressure at the center,
one can obtain the following expressions for the expansion coefficients of the metric function $\Sigma$:
\begin{equation}
\label{bound_coef_ferm}
\Sigma_c=\frac{1}{\sqrt{1/2-B_1 \chi_{2 c}}}, \quad \Sigma_2=\frac{\Sigma_c}{2}\left[1-B_1\left(\chi_{1 c}+\chi_{2 c}\right)\right].
\end{equation}
Again, the sign of $\Sigma_2$ determines whether the system is a single- or double-throat one.

		    As in the case of the polytropic matter,
the dimensional ``kinetic'' energy of the scalar field \eqref{kinet_scal} is determined completely by the characteristic size of the system.
For the fermionic gas under consideration, it is natural to use as a characteristic size the
Landau
radius derived  in considering compact configurations consisting
of an ultrarelativistic degenerate Fermi gas within the framework of Newtonian gravity (see, e.g., Ref.~\cite{Narain:2006kx}):
\begin{equation}
\label{R_Land}
R_L= \frac{\hbar}{c}\frac{M_{\text{Pl}}}{m_f^2},
\end{equation}
where $M_{\text{Pl}}$ is the Planck mass.

In considering our mixed systems with the fermionic fluid, it is then natural to
choose $L=\alpha R_L$, where $\alpha$ is some free scale parameter.
Using this $L$ in the expression for $B_1$ [see after Eq.~\eqref{emt_comp_dmls_fermi}], we get
$B_1=8\pi \alpha^2$. By choosing different values of $\alpha$, we can change
the contribution in the right-hand sides of the Einstein equations  \eqref{emt_comp_dmls_fermi}
coming from the fermionic matter.

\section{Numerical results}
\label{num_calc}

\subsection{Procedure of finding solutions}
\label{num_sol_seek}

When solving the obtained equations numerically, we proceed as in Ref.~\cite{Dzhunushaliev:2014mza},
where the solution search procedure is described in detail. The procedure, briefly, is as follows.
We solve the system of equations \eqref{Einstein-00_dmls}-\eqref{sf_dmls} and \eqref{fluid_dmls}, \eqref{emt_comp_dmls_poly}
(for the polytropic fluid) or \eqref{emt_comp_dmls_fermi}, \eqref{ferm_fluid_dmls} (for the Fermi gas)
with the corresponding boundary conditions \eqref{bound_stat} together with
\eqref{bound_coef} [for the polytropic fluid] or \eqref{bound_chi} and \eqref{bound_coef_ferm} [for the Fermi gas].
In doing so, the configurations under consideration can be subdivided into two regions:
 (i) the internal one, where both the scalar field and the fluid are present;
 (ii) the external one, where only the scalar field is present.
Here the solutions are obtained by using Eqs.~\eqref{Einstein-00_dmls}-\eqref{sf_dmls},
in which $\theta, \chi_1, \chi_2$ are set to zero.

The internal solutions must be matched with the external ones at the boundary of the fluid,
$\xi=\xi_b$, by equating the corresponding values of the functions $\phi$, $\Sigma$, $\nu$
and their derivatives.
The boundary of the fluid $\xi_b$ is defined by $p(\xi_b)=0$.
The value of the integration constant $\nu_c$ at the center is determined
proceeding from the requirement
	       of asymptotic flatness of the external solutions.

As pointed out in Ref.~\cite{Dzhunushaliev:2014mza}, there exists a critical value of $B$, $B_{\text{crit}}$, at which $\Sigma_c\to \infty$ [see Eq.~\eqref{bound_coef}].
Beyond this critical value,
physically reasonable solutions no longer exist. A similar situation takes place for the Fermi gas,
where some critical value  of the coefficient $B_1=B_1^{\text{crit}}$ is also involved, see  Eq.~\eqref{bound_coef_ferm}.
In the present paper we will be interested in solutions corresponding to the values of $B$ and $B_1$
close to the critical ones. Aside from this, as one can see from~\eqref{bound_coef}, as $B\to B_{\text{crit}}$
(that corresponds to $B\sigma \to 1/2$) the expansion coefficient
$\Sigma_2 \sim \left[(1-n)/2 -B\right]$. Then for the polytropic index $n \geq 1$,
which is often used in the literature in modeling relativistic objects,
$\Sigma_2$ will be certainly negative.
That is, if for small values of $B$  there is a single throat
located at the center of the configuration, then, as $B$ increases,
	  the center of the configuration no longer represents a throat
but instead corresponds to an equator. On each side of the equator a minimal area surface (a throat) is located.
In this case the resulting configurations represent double-throat
systems, where the space between the throats
can be completely or partially filled by the fluid.
The latter situation is exactly the one that we consider below.

A similar situation is also found in the case of the Fermi gas.

\subsection{Total mass of the system}

For the spherically symmetric metric \eqref{metric_wh_poten},
the mass $m(r)$ of a volume enclosed by a sphere with
circumferential radius $R_c$, corresponding to the center
		 of the configuration, and another sphere with
circumferential radius $R>R_c$
can be defined as follows:
\begin{equation}
\label{mass_dm}
m(r)=\frac{c^2}{2 G}R_c+\frac{4\pi}{c^2}\int_{R_c}^{r} T_0^0 R^2   dR\equiv
\frac{c^2}{2 G}R_c+\frac{4\pi}{c^2}\int_{0}^{r} T_0^0 R^2 \frac{d R}{d r'}d r' ,
\end{equation}
where we refer to the first term as the mass associated with the center, $M_c$,
while the mass associated with the throat, $M_{\rm th}$, is obtained by
integrating up to the throat radius $R_{\text{th}}=R(r_{\rm th})$.
As pointed out in Ref.~\cite{Dzhunushaliev:2014mza}, despite the fact that
the size of the equator at the center $R_c$ diverges
as $B\to B_{\text{crit}}$,
the size of the throat $R_{\text{th}}$,
and correspondingly the mass associated with the throat, remain finite.
This is because the divergence of
the positive mass 
at the center
$M_c$ is exactly canceled by the mass associated with the mass of the ordinary fluid,
which is negative in the case of double-throat systems.
This comes about
because the derivative  $d R/d r$ is negative in the region where
the ordinary fluid is located, and
therefore the mass integral associated with this fluid gives a negative contribution to the total mass.

In the dimensionless variables
the expression \eqref{mass_dm} takes the form
\begin{equation}
\label{mass_dmls}
m(\xi)=M_*
\left\{\Sigma_c+
\int_{0}^{\xi} \tilde T_0^0 \Sigma^2 \frac{d\Sigma}{d\xi'}d\xi'
\right\},
\end{equation}
where $\tilde T_0^0$ is taken from \eqref{emt_comp_dmls_poly} (for the polytropic fluid) or from \eqref{emt_comp_dmls_fermi}
(for the Fermi gas).
The coefficient $M_*$ in front of the curly brackets has the dimension of mass
$$
M_*^{\text{poly}}=\frac{c^3}{2}\sqrt{\frac{B}{8\pi G^3 \rho_{b c}}} \quad \text{or} \quad
M_*^{\text{Fermi}}=\frac{1}{2}\sqrt{\frac{\hbar^3 c^3 B_1}{8\pi G^3 m_f^4}}
$$
for the polytropic and Fermi fluids, respectively.
Note that the total mass $M$ is then obtained by taking the upper limit of the integral to infinity,
since the energy density of the scalar field becomes equal to zero only asymptotically,
as $\Sigma\to \infty$.

For the case of a massless scalar field considered here, it is useful to write down another,
more elegant definition of the total mass
 via the Komar integral. The latter, in general, is defined as~\cite{Wald}
$$
M_{K}=\frac{2}{c^2}\int_{\Sigma} \left(T_{a b}-\frac{1}{2}g_{a b}T\right)n^a \xi^b dV,
$$
where $n^a$ is a normal to $\Sigma$ and
$\xi^b$ is a timelike Killing vector.
Using the above dimensionless variables,
we find for the polytropic fluid
\begin{equation}
\label{mass_Komar_poly}
M_{K}^{\text{poly}}=M_*^{\text{poly}}B\int_0^{\xi_b}  e^{\nu/2} \Sigma^2\left[1+\sigma(n+3)\theta\right]\theta^n d\xi
\end{equation}
and for the Fermi fluid
\begin{equation}
\label{mass_Komar_Fermi}
M_{K}^{\text{Fermi}}=M_*^{\text{Fermi}}B_1\int_0^{\xi_b}  e^{\nu/2} \Sigma^2\left(\chi_1+3\chi_2\right) d\xi.
\end{equation}
Note that here the integration is performed only in the range from  0 to $\xi_b$, where there is a nonzero contribution
associated with the fluids.

It is seen from the expressions \eqref{mass_Komar_poly} and \eqref{mass_Komar_Fermi} that in order to ensure the finiteness
of the total mass of the system it is necessary that as
$B\to B_{\text{crit}}$ or
$B_1\to B_1^{\text{crit}}$
(when $\Sigma_c\to \infty$)
the metric function $e^{\nu_c}\to 0$ simultaneously
(keeping in mind that the functions $\theta, \chi_1, \chi_2$ remain always finite).
The numerical solutions presented below indicate that in the vicinity of $B_{\text{crit}}$ and $B_1^{\text{crit}}$
this is indeed the case.

\subsection{The choice of the density and pressure at the center}

		      To carry out numerical calculations,
it is necessary to assign the corresponding values of the density and pressure
at the center.
In the vicinity of $B_{\text{crit}}$ and $B_1^{\text{crit}}$,
they can be found from the condition
that the radicand in the expression for  $\Sigma_c$ is approximately equal to zero. Then, taking into account
Eq.~\eqref{B_poly}, for the polytropic EOS we can obtain  from Eq.~\eqref{bound_coef}:
\begin{equation}
\label{rho_bc_crit}
\rho_{bc}=\left[\frac{c^4}{16\pi G L^2 K}(1-\delta)\right]^{n/(n+1)},
\end{equation}
where $\delta\ll 1$ is some constant. In the limit $\delta \to 0$,
the density $\rho_{bc}$ at the center
			  goes to its critical value.

In the same way, for the  Fermi gas, using the condition
that the radicand in the expression for $\Sigma_c$ from \eqref{bound_coef_ferm} is approximately equal to zero,
one can find the value of the pressure at the center
in the vicinity of $B_1^{\text{crit}}$:
\begin{equation}
\label{chi_2c_crit}
\chi_{2c}=\frac{1-\delta}{2 B_1}\equiv \frac{1-\delta}{16 \pi \alpha^2}.
\end{equation}
Here we used the expression for $B_1$ obtained earlier [see after Eq.~\eqref{R_Land}].

Then, using the obtained expressions in the boundary conditions \eqref{bound_coef}, \eqref{bound_stat},
\eqref{bound_chi}, and \eqref{bound_coef_ferm}, we solved the equations numerically  according to the procedure described in Sec.~\ref{num_sol_seek}.

\subsection{Results of calculations}

\begin{figure}[t]
\centering
  \includegraphics[height=8cm]{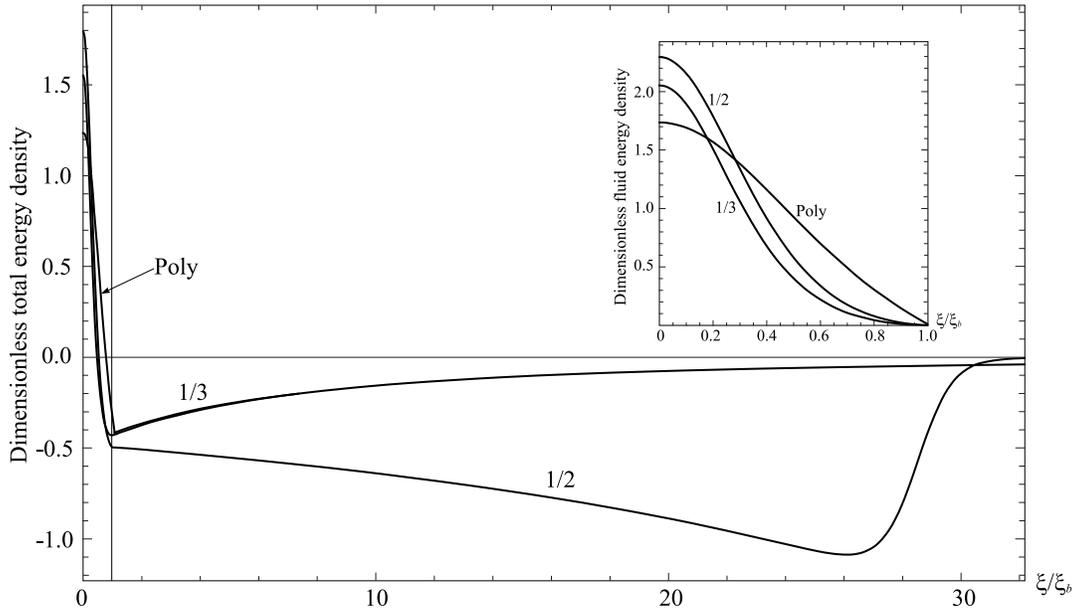}
\caption{The total energy density $\tilde T_0^0$
from Eq.~\eqref{emt_comp_dmls_poly} (for the system with the polytropic fluid) and from Eq.~\eqref{emt_comp_dmls_fermi}
(for the system with the fermionic fluid)
(both in units of $\varphi_1^2$) are shown as functions of the relative radius $\xi/\xi_b$.
The inset shows
the fluid energy densities $B  (1+\sigma n \theta) \theta^n$ (for the polytropic fluid)
and $B_1  \chi_1$ (for the fermionic fluid). The numbers near the curves correspond to the values of the scale parameter
 $\alpha$ for the Fermi systems.
 Since the solutions are symmetric with respect to $\xi=0$, the graphs are shown only
for $\xi>0$.  The thin vertical line corresponds to the boundary of the fluids.
For all plots, the parameter $\delta$ is taken as $10^{-15}$.
Asymptotically, as $\xi\to\pm\infty$, the total energy goes to zero for all systems.
}
\label{fig_energ_mix}
\end{figure}

\begin{figure}[h!]
\centering
  \includegraphics[height=10cm]{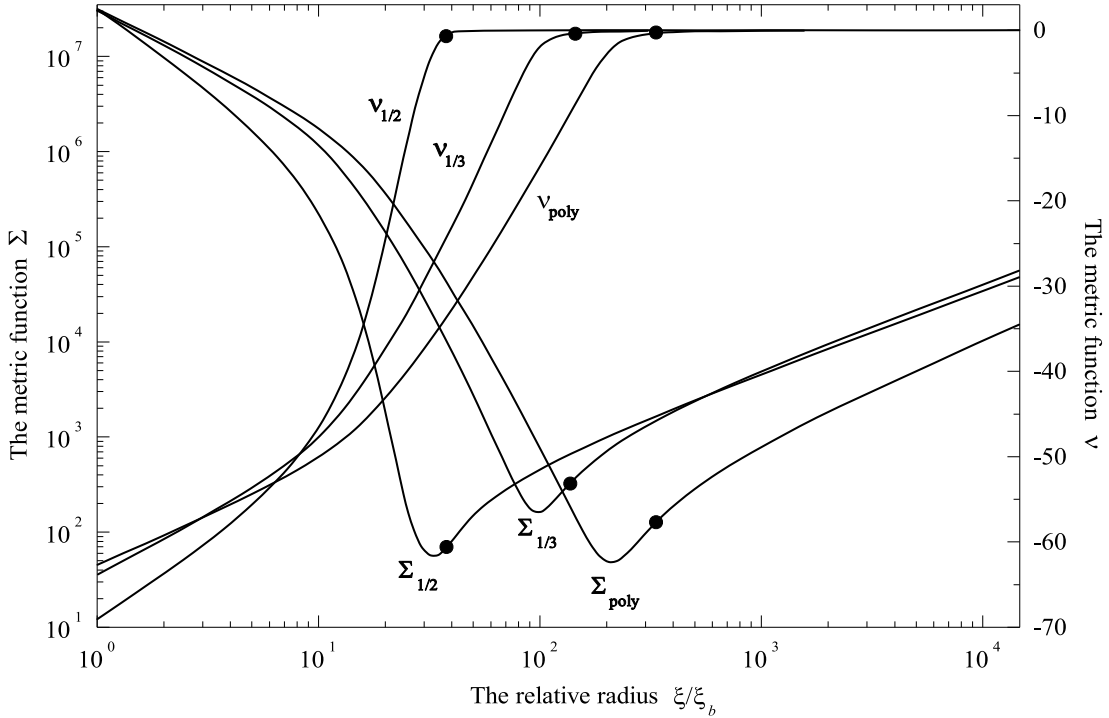}
\caption{External metric functions $\Sigma$ and $\nu$ for the systems
of Fig.~\ref{fig_energ_mix} are shown as functions of the relative radius $\xi/\xi_b$.
The minima of the function $\Sigma$ correspond to the location of the throat.
The bold dots denote the minimum radius for a stable orbit $\xi_{\text{so}}$ (see below in Sec.~\ref{sec_thin_acc_disk}).
Asymptotically, as $\xi\to\pm\infty$, the spacetime is flat
with $\Sigma\to |\xi|$ and $e^{\nu}\to 1$
from  below.
}
\label{fig_sigma_nu_exter}
\end{figure}

\begin{figure}[t]
\centering
  \includegraphics[height=8cm]{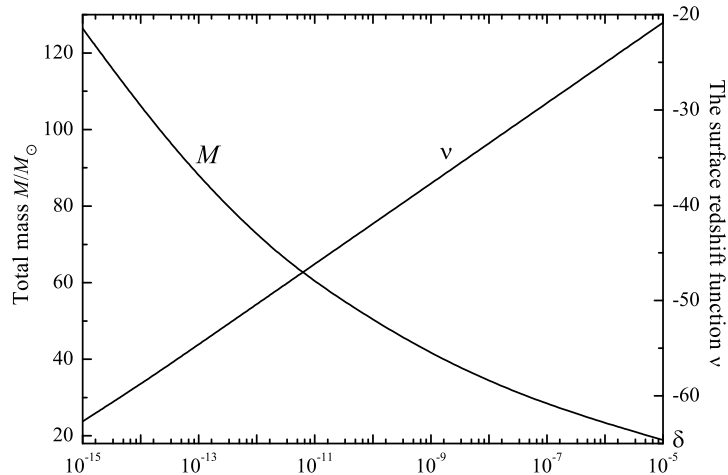}
\vspace{-1.2cm}
\caption{Total mass and the surface value of the redshift function $\nu^{\text{surf}}_{\text{poly}}$
for the system with the polytropic fluid are shown as functions of the parameter
$\delta$. For all systems, the radius of the fluid is
$R_{\text{poly}}\approx 10.55\, \text{km}$.
}
\label{fig_mass_nu_poly}
\end{figure}

\begin{figure}[h!]
\begin{minipage}[t]{.5\linewidth}
  \begin{center}
  \includegraphics[width=9cm]{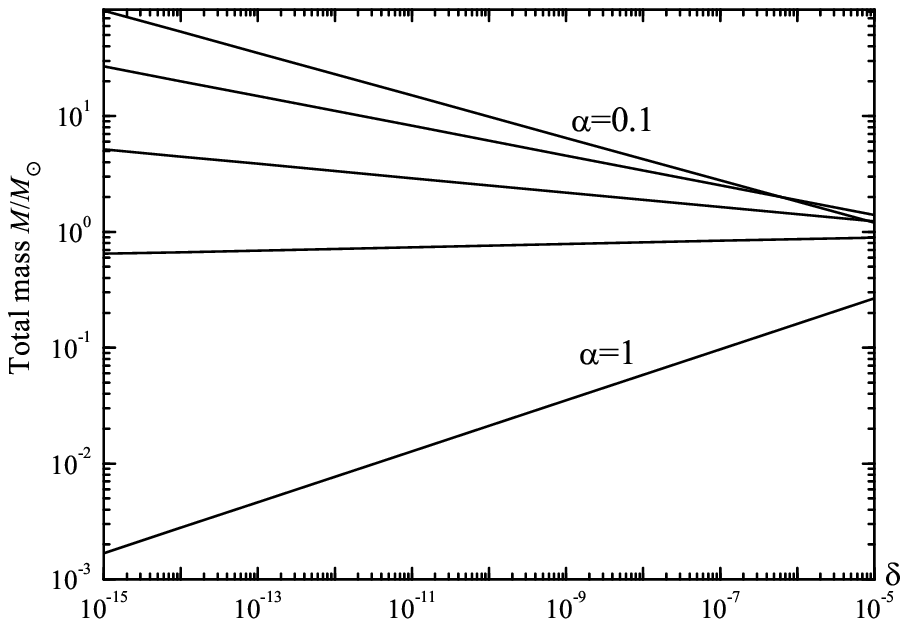}
  \end{center}
\end{minipage}\hfill
\begin{minipage}[t]{.5\linewidth}
  \begin{center}
  \includegraphics[width=9cm]{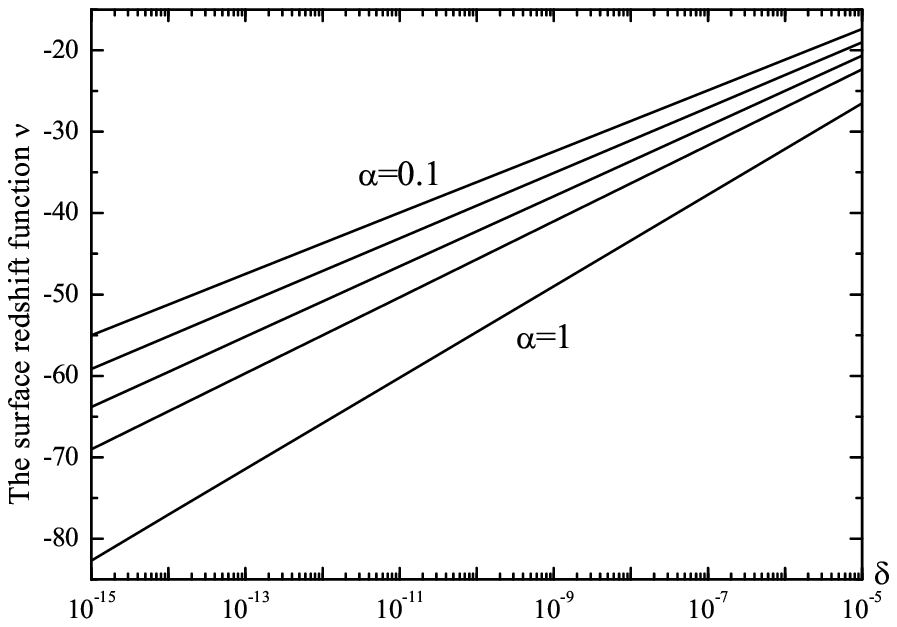}
  \end{center}
\end{minipage}\hfill
\vspace{-.5cm}
  \caption{Total mass and the surface value of the redshift function $\nu^{\text{surf}}_{\text{Fermi}}$ for the systems with the Fermi fluid
are shown as functions of the parameter $\delta$.
For all systems,  the fermion mass is taken to be $m_f=1\, \text{GeV}$.
   The parameter $\alpha$
takes the values $1/10, 1/5, 1/3, 1/2, 1$,
from top to bottom.
The radii of the fluid are:
for $\alpha=1$ -- $R_{\text{Fermi}}\approx 2.08\, \text{km}$;
for $\alpha=1/2$ -- $R_{\text{Fermi}}\approx 1.44\, \text{km}$;
for $\alpha=1/3$ -- $R_{\text{Fermi}}\approx 1.14\, \text{km}$;
for $\alpha=1/5$ -- $R_{\text{Fermi}}\approx 0.84\, \text{km}$;
for $\alpha=1/10$ -- $R_{\text{Fermi}}\approx 0.55\, \text{km}$.
  }
 \label{fig_mass_nu_fermi}
\end{figure}

Examples of the obtained solutions are presented in Figs.~\ref{fig_energ_mix}-\ref{fig_mass_nu_fermi}.
Fig.~\ref{fig_energ_mix} shows the typical distributions of the total and fluid
energy densities for the mixed systems under consideration.
The calculations indicate that when
the scale parameter $\alpha \gtrsim 1/2$
the total energy density has a characteristic well in the neighborhood of the throat located at
 $\xi=\xi_{\text{th}}$. This is because the
``kinetic'' energy of the scalar field, which behaves here as
$\sim e^{\nu_c-\nu_{\text{th}}}\left(\Sigma_c/\Sigma_{\text{th}}\right)^4$
[see Eq.~\eqref{sf_dmls}], exhibits fast growth due to the large value of
$\Sigma_c$, on one hand, and to the small value of
$\Sigma_{\text{th}}$, on the other hand. This is accompanied by a simultaneous fast decrease (modulus)
of the metric function $\nu$, which, starting from large values on the surface of the fluid,
undergoes a sharp decrease as the throat is approached
(see Fig.~\ref{fig_sigma_nu_exter}). The ultimate result is that, for instance, at
 $\alpha=1$ in the vicinity of $\xi=\xi_{\text{th}}$ the
dimensionless total energy density is of the order of $|4\times 10^{5}|$ (at $\delta=10^{-15}$).

Fig.~\ref{fig_sigma_nu_exter} shows the typical behavior of the metric functions
 $\Sigma$ and $\nu$ in the external region of the mixed systems. For the configurations under consideration
 with $B\approx B_{\text{crit}}$ and $B_1\approx B_1^{\text{crit}}$,
the throats always reside outside the fluid
 (for systems with the throats located within a fluid, see Ref.~\cite{Dzhunushaliev:2014mza}).
The numerical calculations indicate that
 as $\alpha$ decreases, the throat shifts further away from the surface of the fluid
(in units of relative radii).

Figs.~\ref{fig_mass_nu_poly} and \ref{fig_mass_nu_fermi} show the total masses and the values of the metric function
$\nu$ on the surface of the fluid,
 $\nu^{\text{surf}}_{\text{poly}}=\nu(\xi_b)$
 and $\nu^{\text{surf}}_{\text{Fermi}}=\nu(x_b)$,
as functions of the parameter $\delta$.
As $\delta \to 0$, the mass of the polytropic configurations increases, tending to some finite value.
This growth of the mass is accompanied by a simultaneous increase of
the modulus of $\nu^{\text{surf}}$.
Thus, from the point of view of a distant observer,
the surface of the fluid will look like a high-redshift surface, i.e., in this sense the system will be similar to a BH.
In turn, for any $\delta$ used here the radii of the fluids remain almost unchanged
(for the numerical values, see in captions of Figs.~\ref{fig_mass_nu_poly} and~\ref{fig_mass_nu_fermi}),
but the gravitational radius of the systems,
$r_g=2.95 (M/M_{\odot}) \,\text{km}$, grows as $\delta \to 0$. Note that for the configurations considered here
the radius of the polytropic fluid is always less than the gravitational radius of the system as a whole.

A similar situation also takes place for the system with the fermionic fluid. The only difference is in the behavior
of the mass of the configuration for large $\alpha$,
when as $\delta \to 0$ the mass does not grow, but decreases
(see the left panel of Fig.~\ref{fig_mass_nu_fermi}).
There, as in the case of the polytropic fluid,
the modulus of $\nu^{\text{surf}}_{\text{Fermi}}$ increases for any
$\alpha$ (see the right panel of Fig.~\ref{fig_mass_nu_fermi}).
In turn, the radius of the Fermi fluid is always
(for any $\delta$) larger than the gravitational radius for
$\alpha=1$ and less than it for $\alpha \leq 1/2$.

The numerical values of the masses and sizes of the systems with the Fermi fluid shown in Fig.~\ref{fig_mass_nu_fermi}
are given for $m_f=1\, \text{GeV}$. For other  $m_f$
the values of the total masses and sizes are derived from those of Fig.~\ref{fig_mass_nu_fermi}
by multiplying them by the factor $\left(1\,\text{GeV}/m_f\right)^2$, where
 $m_f$ is taken in GeV.
There, the value of
$\nu^{\text{surf}}_{\text{Fermi}}$ will remain the same for any $m_f$. Correspondingly,
if we assume for definiteness that $m_f$ lies in the range $1~\text{eV} \lesssim m_f \lesssim 10^{2}~\text{GeV}$
(such values are used, for instance, in modeling dark matter~\cite{Narain:2006kx}), the total masses and sizes of the configurations under consideration
will lie in a very wide range. For example, for $m_f$ equal, say, to $1\,\text{MeV}$, the system
with $\alpha=1/3$ and $\delta=10^{-15}$ has the total mass
$M\sim5\times 10^6 M_\odot$, and the radius of the surface of the fluid is
 $R\sim 10^6\, \text{km}$. A configuration with such characteristics, possessing a high-redshift surface, might mimic BHs at the center
 of galaxies~\cite{Shodel:2002}.

\section{Thin accretion disk}
\label{sec_thin_acc_disk}

In this section we consider the process of accretion of test particles onto our configurations.
The purpose is to find out what are the differences between the mixed systems under consideration and BHs
as regards the observational manifestations associated with the accretion process.
To do this, we consider a relativistic model of a thin accretion disk and analyze  the properties of the energy flux
emitted from the disk's surface.
Let us emphasize that here we do not consider the question of the infall of accreting matter (plasma) onto the surface
of the fluid and of changes in the emission spectra associated with such a process, but
restrict ourselves to the consideration of phenomena related to the accretion disk only.

We will closely follow the work of Page and Thorne~\cite{Page:1974he}, who studied the model
of thin-disk accretion onto a BH. In considering the accretion process, it is assumed
that the accretion disk consists of gas particles rapidly orbiting a central strongly gravitating body.
In such a process, the gas slowly
loses angular momentum, for example, due to the magnetic and/or turbulent viscosity~\cite{Shakura:1972te,Novikov:1973,Shapiro:2008}.
It causes the gas to move slowly inwards, losing gravitational potential energy and
heating up the accretion disk.
Eventually the gas has lost enough angular momentum that it can no longer
follow a stable circular orbit, and so it spirals rapidly inwards onto the central
object.
A fraction of the heat converts into radiation and cools down the disk.
The resulting emission may lie in various wavelength ranges (radio, optical, and X-ray),
and the analysis of its spectrum enables one to reveal the distinguishing features of objects onto which the accretion occurs.

The model of Ref.~\cite{Page:1974he} is based on a number of assumptions, including:
(i) It is assumed that the accretion disk has negligible self-gravity and reside in an external spacetime geometry
(BH geometry in~\cite{Page:1974he}). (ii) The disk lies in the equatorial plane of the central object.
(iii) The disk is assumed to be thin, i.e., its thickness is much less than its radius. (iv) The physical quantities describing the gas
in the disk are averaged over a characteristic time interval $\Delta t$ and the azimuthal angle $\Delta \varphi=2\pi$
(not to be confused with the scalar field). (v) The energy-momentum tensor of disk matter may contain any types of stress-energy.
(vi) Heat flow within the disk is assumed to be negligible, except in the vertical direction.

Based on these assumptions and using the laws of conservation of rest mass, angular momentum, and energy,
one can derive the following expression for the time-averaged flux of radiant energy flowing out of the upper or lower side of the disk~\cite{Page:1974he}:
\begin{equation}
\label{flux}
F(r)=-\frac{\dot{M}_0 c^2}{4\pi \sqrt{-g}}\frac{\Omega_{,r}}{\left(\bar{E}-\Omega\bar{L}\right)^2}\int_{r_{\text{ISCO}}}^{r}
\left(\bar{E}-\Omega\bar{L}\right)\bar{L}_{,r} dr.
\end{equation}
Here $\Omega$, $\bar{L}$, and $\bar{E}$ are the angular velocity, the specific angular momentum, and the
specific energy of particles moving in circular orbits around the central body, respectively;
$\dot{M}_0=\text{const.}$ is the time-averaged rate at which rest mass flows inward through the disk.
The subscript $,r$ denotes differentiation with respect to $r$.
The lower limit of integration $r_{\text{ISCO}}$ corresponds to the innermost stable
 circular orbit (ISCO) from which
the accreting matter starts to fall freely onto the central object. In the derivation of formula \eqref{flux}
it is assumed
that the ``no torque'' condition, according to which the torque vanishes at the inner edge of the disk, is satisfied~\cite{Page:1974he}.
Such a condition is only valid when no strong magnetic fields exist in the plunging region.

All quantities entering Eq.~\eqref{flux} depend on the
radial coordinate $r$ only. According to the above assumptions (ii) and (iii),
to describe the accretion process, one can
introduce the following cylindrical line element in and near the equatorial plane ($|\theta-\pi/2|\ll 1$):
\begin{equation}
\label{metr_equator}
ds^2=e^{2\gamma}(d x^0)^2-e^{2\alpha}dr^2-e^{2\beta}d\varphi^2-dZ^2,
\end{equation}
where $\alpha, \beta, \gamma$ are functions of $r$ only.
[This metric is derived from the general spherically symmetric one by replacing the usual angular coordinate $\theta$
by $Z=e^\beta\cos{\theta}\approx e^\beta(\theta-\pi/2)$.]

Using this metric, let us integrate the geodesic equation. Since here we consider
timelike geodesics for massive particles, one can obtain the following expressions for
the specific energy and the specific angular momentum: $\bar{E}=c^2 e^{2\gamma} \dot{t}$ and
$\bar{L}=e^{2\beta}\dot\varphi$, where the dot denotes  differentiation with respect to the proper time $\tau$
along the path.

Next, using a first integral of the geodesics equations $g_{\mu\nu}\dot{x}^\mu\dot{x}^\nu=c^2$ and substituting into it
the above expressions for  $\bar{E}$ and  $\bar{L}$, one can find the following ``energy'' equation for a particle
\begin{equation}
\label{energ_cons_part}
\frac{\bar{E}^2}{c^2}=e^{2(\alpha+\gamma)}\dot{r}^2+V_{\text{eff}},
\end{equation}
where the effective potential is given by
\begin{equation}
\label{eff_poten}
V_{\text{eff}}(r)=e^{2\gamma}\left(c^2+e^{-2\beta} \bar{L}^2\right).
\end{equation}

 For circular motion in the equatorial plane, we obviously have $r=\text{const.}$ Correspondingly, it follows from \eqref{energ_cons_part}
that $V_{\text{eff},r}=0$. Using this condition together with Eq.~\eqref{energ_cons_part} and the definition of the angular velocity
$\Omega=d\varphi/dt$, one can obtain the following expressions
\begin{eqnarray}
\label{ang_vel}
&&\Omega=c e^{\gamma-\beta}\sqrt{\frac{\gamma_{,r}}{\beta_{,r}}},\\
\label{ang_mom}
&&\bar{L}=\frac{c \Omega e^{2\beta}}{\sqrt{c^2 e^{2\gamma}-e^{2\beta}\Omega^2}},\\
\label{energ_specif}
&&\bar{E}=\frac{c^3 e^{2\gamma}}{\sqrt{c^2 e^{2\gamma}-e^{2\beta}\Omega^2}}.
\end{eqnarray}
Using them in \eqref{flux}, one can find a dependence
of the energy flux on the radius.

Since here we consider the mixed configurations closely mimicking black hole
features, it is interesting to compare the resulting fluxes for Schwarzschild BHs
 (SBHs) and our systems. To do this, let us first rewrite the obtained expressions in terms of dimensionless variables.

\subsection{The case of  Schwarzschild black holes}

The characteristic size of a SBH is its gravitational radius $r_g$. Hence, it is convenient to introduce the dimensionless radius
$x=\left(c^2/G M\right)r\equiv 2 r/r_g$, where
$M$ is the mass of the SBH. Then, taking into account that for this case the metric functions from
\eqref{metr_equator} are
$$
e^{2\gamma}=e^{-2\alpha}=1-\frac{r_g}{r}, \quad e^{2\beta}=r^2,
$$
we have from \eqref{ang_vel}-\eqref{energ_specif}:
 \begin{equation}
\label{func_for_BH}
\Omega_{\text{SBH}}=\frac{2c}{r_g} x^{-3/2},\quad
\bar{L}_{\text{SBH}}=\frac{c \,r_g}{2}\frac{x}{\sqrt{x-3}}, \quad
\bar{E}_{\text{SBH}}=c^2\left(1-\frac{2}{x}\right)\sqrt{\frac{x}{x-3}}.
\end{equation}

As is well known (see, e.g., Ref.~\cite{Zeld}), for a SBH, the radius
$x=3$ (or $r=3 r_g/2$) corresponds to the radius for an unstable circular orbit.
In turn, for the ISCO $x_{\text{ISCO}}=6$ (or $r_{\text{ISCO}}=3 r_g$). This orbit is
 unique in satisfying both $V_{\text{eff},r}=0$ and $V_{\text{eff},r r}=0$.
 The free radial infall of particles
of the accretion disk starts from this orbit. Therefore, the lower limit of integration in
\eqref{flux} is $r_{\text{ISCO}}$.

Taking all this into account, and also that for the metric \eqref{metr_equator} $\sqrt{-g}=r$,
one obtains the following expression for the flux \eqref{flux}:
 \begin{equation}
\label{flux_BH}
F_{\text{SBH}}(x)=\frac{3}{2}\frac{\dot{M}_0 c^2}{\pi r_g^2}\frac{1}{x^{5/2}(x-3)}\left\{\sqrt{x}-\sqrt{6}+\frac{\sqrt{3}}{2}
\left[\ln{\left(\frac{\sqrt{x}+\sqrt{3}}{\sqrt{6}+\sqrt{3}}\right)}-
\ln{\left(\frac{\sqrt{x}-\sqrt{3}}{\sqrt{6}-\sqrt{3}}\right)}\right]
\right\}.
\end{equation}

\subsection{The case of cold black holes}

Another type of objects possessing an event horizon (both regular and singular) is the so-called
cold black holes (CBHs)~\cite{Bronnikov:1998gf,Bronnikov:1998hm,Bronnikov:2011if,Bronnikov:2006qj}.
They can be obtained, in particular, as a solution of the Einstein equations with the matter source
in the form of a massless scalar field~\cite{Bronnikov:2011if} or the field
with a potential energy~\cite{Bronnikov:2006qj}. Their distinctive feature is the presence
of the horizon of infinite area and correspondingly vanishing Hawking temperature.

The line element of CBHs is given by~\cite{Bronnikov:2011if}
\begin{equation}
ds^2 = c^2 p^a dt^2 - p^{-a} dr^2 - r^2 p^{1-a} d\Omega^2 \ , \
\label{met_cbh}
\end{equation}
with $p =1 -r_0/r$, where $r_0$ and the gravitational radius $r_g$
are related by $r_0=r_g/a$.

In the special case $a=2$ the CBHs possess a regular event horizon
on which the curvature invariants remain finite
(for a detailed discussion, see Refs.~\cite{Bronnikov:1998gf,Bronnikov:1998hm,Bronnikov:2011if}).

In order to compute the energy flux of radiant energy of the disk for such a CBH,
we set
$$
e^{2\gamma}=e^{-2\alpha}=\left(1-\frac{r_0}{r}\right)^2,
      \quad  e^{2\beta}\equiv\mathcal{R}^2=\frac{r^2}{1-\frac{r_0}{r}} \ .
$$
Then, using Eqs.~\eqref{ang_vel}-\eqref{energ_specif},
we find
 \begin{equation}
\label{func_for_CBH}
\Omega_{\text{CBH}}=\sqrt{2}\frac{c}{r_0}\left[\frac{(z-1)^3}{z^5(2z-3)}\right]^{1/2},\quad
\bar{L}_{\text{CBH}}=c \sqrt{2} r_0
\left[\frac{z^3}{(2z-5)(z-1)}\right]^{1/2}, \quad
\bar{E}_{\text{CBH}}=c^2 \frac{z-1}{z}
\left[\frac{2z-3}{2z-5}\right]^{1/2}\ ,
\end{equation}
with $z=r/r_0$,
and for the radial coordinate of the ISCO
$z_{\text{ISCO}} = (7+\sqrt{19})/2$.
Then the
circumferential radius $\mathcal{R}_{\text{ISCO}}$
(scaled by the gravitational radius) of the ISCO is equal to
$z_{\text{ISCO}}/(2\sqrt{1 -1/z_{\text{ISCO}}})\approx 3.128$.

In principle, the flux $F_{\text{CBH}}(z)$ can be expressed in terms of elliptic functions.
However, the expression is not very instructive. Therefore, we omit it here.

\subsection{The case of mixed system}

The characteristic size of the mixed systems is $L$ from \eqref{dimless_xi_v}.
According to Eq.~\eqref{metric_wh_poten}, the metric functions appearing in
\eqref{metr_equator} are
 $\gamma=\nu/2, \alpha=0, e^\beta=R$. Then we have from Eqs.~\eqref{ang_vel}-\eqref{energ_specif}:
 \begin{equation}
\label{func_for_mixed}
\Omega_{\text{mix}}=\frac{c e^{\nu/2}}{L} \sqrt{\frac{\nu'}{2\Sigma \Sigma'}},\quad
\bar{L}_{\text{mix}}=c \,L\sqrt{\frac{\Sigma^3 \nu'}{2\Sigma'-\Sigma \nu'}}, \quad
\bar{E}_{\text{mix}}=c^2\sqrt{\frac{2 e^\nu \Sigma' }{2\Sigma'-\Sigma \nu'}}.
\end{equation}
Here the prime denotes differentiation with respect to $\xi$ from  \eqref{dimless_xi_v}.
Substituting these expressions into Eq.~\eqref{flux}, one can calculate the corresponding flux for the mixed systems under consideration:
 \begin{equation}
\label{flux_mixed}
F_{\text{mix}}(\xi)=-\frac{\dot{M}_0 c^2}{4\pi L^2}\frac{\Omega_{\text{mix}}^\prime}{e^{\nu/2}\Sigma\left(\bar{E}_{\text{mix}}-
\Omega_{\text{mix}}\bar{L}_{\text{mix}}\right)^2}\int_{\xi_{\text{ISCO}}}^{\xi}
\left(\bar{E}_{\text{mix}}-\Omega_{\text{mix}}\bar{L}_{\text{mix}}\right)\bar{L}_{\text{mix}}^\prime d\xi.
\end{equation}
Note that  $\Omega_{\text{mix}},\bar{L}_{\text{mix}}$, and $\bar{E}_{\text{mix}}$ entering Eq.~\eqref{flux_mixed}
are taken from \eqref{func_for_mixed} without the dimensional coefficients, i.e., without  $c$ and $L$.

In turn, the effective potential \eqref{eff_poten} takes the form:
\begin{equation}
\label{eff_poten_mixed}
V_{\text{eff}}^{\text{mix}}(\xi)=c^2\frac{2 e^\nu \Sigma' }{2\Sigma'-\Sigma \nu'}.
\end{equation}

Taking into account that for the systems under consideration the positive function
$\Sigma$ does have a minimum somewhere in the external region, i.e., at
$\xi>\xi_b$, and $\nu'$ always exceeds zero, it is seen from
Eqs.~\eqref{func_for_mixed} and \eqref{eff_poten_mixed} that the denominator
$(2\Sigma'-\Sigma \nu')$ inevitably crosses zero  somewhere. Then
$\bar{E}\to \infty, \bar{L}\to \infty$ at that point, and, analogously to a BH, this point corresponds to the minimum radius for an unstable orbit.
It may also be noted that, since $\Sigma'>0$ only in the region lying outside the throat
(i.e., at $\xi>\xi_{\text{th}}$), all circular orbits (and correspondingly the accretion disk) will certainly lie
outside the throat (see Fig.~\ref{fig_sigma_nu_exter} where the coordinates of the
ISCOs
$\xi_{\text{ISCO}}$ are shown by bold dots). Note also that, since asymptotically $\Sigma\to \xi$ and $\nu^\prime\sim \xi^{-2}$,
 $(2\Sigma'-\Sigma \nu')\to 2$.

Next, the remaining circular orbits can be found from the condition $d V_{\text{eff}}^{\text{mix}}/d\xi=0$,
and the orbits are stable or unstable if
 $d^2 V_{\text{eff}}^{\text{mix}}/d\xi^2>0$ or $d^2 V_{\text{eff}}^{\text{mix}}/d\xi^2<0$, respectively.
Notice that  $V_{\text{eff}}^{\text{mix}}/c^2\to 1$ asymptotically.

\subsection{Results of calculations}

Using Eq.~\eqref{flux_BH}, one can calculate
the maximum value of the flux for a SBH:
\begin{equation}
\label{BF_max_flux}
F_{\text{SBH}}^\text{max}=1.72\times 10^{-4}\frac{\dot{M}_0 c^2}{\pi r_g^2}\,\text{erg}\,\text{cm}^{-2}\,\text{sec}^{-1}.
\end{equation}
This maximum value is always (whatever the black hole's mass) reached at the same radius
$x_\text{max}=9.55$,
or, in dimensional units, at
$r_\text{max}=4.78\, r_g$.

Then, to compare the fluxes of the mixed systems and SBHs with the same masses, it is convenient to express the flux \eqref{flux_mixed}
in units of $F_{\text{SBH}}^\text{max}$. This yields
\begin{equation}
\label{flux_mixed_in_FBH}
\frac{F_{\text{mix}}}{F_{\text{SBH}}^\text{max}}=-1.45\times 10^3\left(\frac{r_g}{L}\right)^2
\frac{\Omega_{\text{mix}}^\prime}{e^{\nu/2}\Sigma\left(\bar{E}_{\text{mix}}-
\Omega_{\text{mix}}\bar{L}_{\text{mix}}\right)^2}\int_{\xi_{\text{ISCO}}}^{\xi}
\left(\bar{E}_{\text{mix}}-\Omega_{\text{mix}}\bar{L}_{\text{mix}}\right)\bar{L}_{\text{mix}}^\prime d\xi.
\end{equation}

Similarly, we scale the flux of the CBH by $F_{\text{SBH}}$.

\begin{figure}[t]
\centering
  \includegraphics[height=10cm]{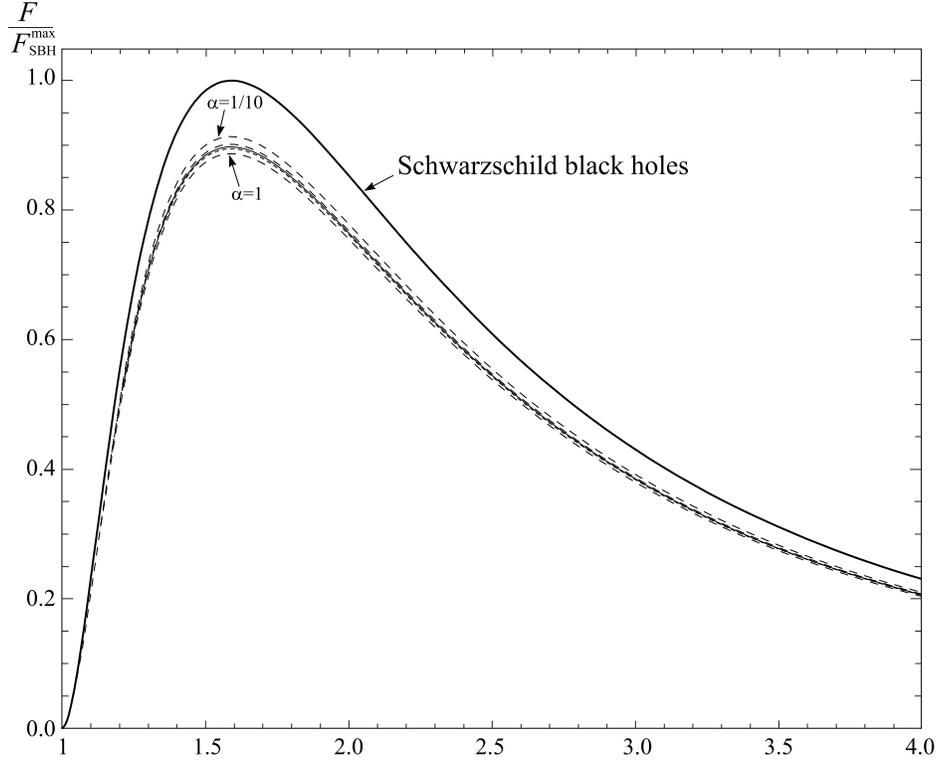}
\caption{The fluxes for the mixed systems and BHs expressed in units of $F_{\text{SBH}}^\text{max}$ from \eqref{BF_max_flux}.
The graphs for the system with the polytropic and Fermi fluids are shown
by thin solid and long-dashed lines, respectively.
The Schwarzschild black holes and cold black holes are represented
by the thick solid line and the short-dashed line, respectively.
The curves for the  polytropic case and the cold black hole (nearly) coincide.
The abscissa indicates the sizes of the disk in units of $x_{\text{ISCO}}$ (for a SBH) and $\Sigma_{\text{ISCO}}$ (for the mixed systems).
}
\label{fig_fluxes}
\end{figure}

\begin{figure}[h!]
\centering
  \includegraphics[height=10.cm]{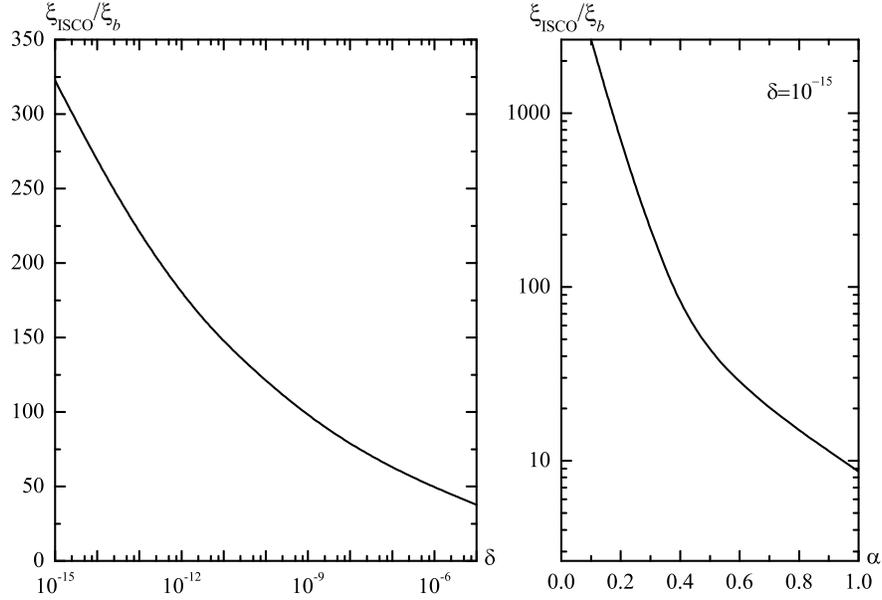}
\vspace{-1.4cm}
\caption{The proper distance between the ISCO and the center (measured in units of the proper distance between the fluid surface and the center)
for the mixed systems with the polytropic (left panel) and Fermi (right panel) fluids.
}
\label{fig_min_stab_orb}
\end{figure}

The corresponding graphs are plotted in Fig.~\ref{fig_fluxes}.
For purposes of comparison, it seems more informative to work in relative units where the circumferential radius of the mixed systems
$\Sigma$ is normalized to the radius for the ISCO, $\Sigma_{\text{ISCO}}$
(for some mixed systems the location of the stable orbit is shown in Fig.~\ref{fig_sigma_nu_exter} by bold dots),
and the circumferential radius of SBHs $x$ is analogously normalized to $x_{\text{ISCO}}$.
Then the flux for the mixed systems is calculated from formula
\eqref{flux_mixed_in_FBH}, and for SBHs -- from \eqref{flux_BH}, both in units of \eqref{BF_max_flux}.
Also, the figure shows the flux for the CBH.
In this case the profile of the flux distribution for SBHs remains unchanged for any mass, in contrast to the profile
of the mixed systems which changes depending on the particular values of the system parameters. Notice, however, that for the mixed system with the polytropic EOS
the dependence on $\delta$ is too small to be seen in the graph.
Moreover, the graphs for the CBH and the mixed system for $\delta=
10^{-15}$ nearly coincide. Thus, we conclude that
in the limit of vanishing $ \delta$ (i.e., when $B\to B_{\text{crit}}$) the solution outside the
neutron star coincides with the CBH solution.
For the mixed system with the Fermi gas the fluxes for different
values of $\alpha$ can be distinguished, although they are close
to the CBH case.

Compared with the case of a SBH, it is characteristic for the fluxes of the mixed system that
the maximum value is always lower,  reaching
 $\sim 90\%$ of the maximum for SBHs in all cases.
In any case, for both types of mixed systems
the distribution of the flux along the radius differs considerably from the case of SBHs with the same masses.

Note also one more difference between the mixed systems under consideration
and SBHs concerning the location of the ISCO.
In terms of the circumferential radius the ISCOs of a SBH
are always located at $r_{\rm ISCO}=3 r_g$  and those of a CBH at
$\mathcal{R}_{\text{ISCO}}\approx 3.128\, r_g$.
For the mixed system we find that for the polytropic EOS
$R_{\,\text{ISCO}} \approx 3.12\, r_g$, almost independent of the parameter $\delta$,
while for the fermionic EOS $R_{\,\text{ISCO}}$ varies between $\approx 3.10 r_g$
for $\alpha=0.1$ and
$\approx 3.13 r_g$ for $\alpha=1.0$.

Since for wormholes the circumferential radius is not an appropriate
quantity to specify the location of the ISCO, we present in Fig.~\ref{fig_min_stab_orb} the proper distance between the
ISCO and the center in units of the proper distance between the
fluid boundary and the center. We note that for the mixed systems this
quantity strongly depends on the parameters $\delta$ and $\alpha$.

Let us now  estimate the efficiency of energy radiation, $\epsilon$, in an accretion disc.
The maximum efficiency is of the order of the ``gravitational binding energy''
at the ISCO (i.e., the energy lost by a particle due to motion
from infinity to the lowest orbit) divided by the rest mass energy of the particle.
For a SBH, the lowest orbit is the ISCO with $x_{\text{ISCO}}=6$,
for a CBH -- $z_{\text{ISCO}} = (7+\sqrt{19})/2$, and
for the mixed system -- $\Sigma_{\text{ISCO}}$. Then, using the expressions for the specific energy
\eqref{func_for_BH}, \eqref{func_for_CBH}, and \eqref{func_for_mixed},
we find that for the configurations considered above (with values of the parameters
$\alpha$ and $\delta$ for which the graphs of Figs.~\ref{fig_mass_nu_poly}-\ref{fig_min_stab_orb} are plotted) the efficiency is:
$$
\epsilon_{\text{SBH}}=1-\bar{E}_{\text{SBH}}/c^2\approx 0.05719, \quad \epsilon_{\text{CBH}}=1-\bar{E}_{\text{CBH}}/c^2\approx 0.05535, \quad \epsilon_{\text{mix}}=1-\bar{E}_{\text{mix}}/c^2\approx 0.055.
$$
That is, the efficiency of the conversion of the accreted mass into radiation
for the CBH and the mixed configurations under consideration is approximately the same  ($\sim 5.5\%$) and differs slightly from the SBHs $\sim 5.7\%$.

Consider now the question of the spectrum emitted from the disk's surface. For this purpose,
we must determine the spectrum
emitted locally at each point of the disk and then integrate over the whole disk surface. Starting from the assumption
that the disk is optically thick, in the sense that each element of the disk radiates as a black
body with temperature $T(r)$, one can define this temperature via the above flux by using formula
$F(r)=\sigma_{\text{SB}} T^4(r)$,
where $\sigma_{\text{SB}}$ is the Stefan-Boltzmann constant. The total energy of such radiation (from both sides of the disk)
 at frequency  $\omega$ is
$$
S(\omega)=2 \int I(\omega) d S_d
\quad
\text{with}  \quad I(\omega)=\frac{\hbar \omega^3}{2\pi^2 c^2}\frac{1}{e^{\hbar\omega/k_B T}-1},$$
where $I(\omega)$ is the Planck distribution function,
$k_B$ is the Boltzmann constant. The surface area of the disk
$S_d$ is: $$S_d=2\pi\int_{r_{in}}^{r_{out}}e^{\beta} dr,$$
where $r_{in}$ and $r_{out}$ are the inner and outer radii of the disk
 [recall that $\beta$ is the metric function from \eqref{metr_equator}].
(Note that in calculating the above energy it is assumed that the radiation is formed on the disk surface at a depth of
$\tau_{\text{ff}}\approx 1$~\cite{Shakura:1972te}, where the subscript ``$\text{ff}$'' refers to free-free emission.)

Using these expressions, one can obtain:
\begin{equation}
\label{energ_disk_gen}
S(\omega)=\frac{2\hbar}{c^2}\omega^3 \int_{r_{in}}^{r_{out}}\frac{e^{\beta}}{e^{\hbar\omega/k_B T}-1}dr.
\end{equation}
If the disk is inclined with respect to an observer at angle $i$, defined as the angle between the line of sight and the normal to the disk,
then, to calculate the measured energy, the above expression should be multiplied by $\cos{i}$.

Since in the present paper our aim is to reveal the observational differences between BHs and mixed systems,
it is more informative to calculate the relative energy of the mixed system $S_{\text{mix}}$
expressed in terms of the radiant energy $S_{\text{SBH}}$ corresponding to a SBH with the same mass.
In the above dimensionless variables this relation takes the form:
\begin{equation}
\label{energ_disk_relative}
\frac{S_{\text{mix}}(\omega)}{S_{\text{SBH}}(\omega)}=4\left(\frac{L}{r_g}\right)^2 \int_{\xi_{\text{ISCO}}}^{\xi_{out}} \frac{\Sigma d\xi}{e^{\hbar\omega/k_B T}-1}\Big{/}
\int_{x_{\text{ISCO}}}^{x_{out}} \frac{x d x}{e^{\hbar\omega/k_B T}-1}.
\end{equation}
Since formally the disk extends to infinity, the upper limit of the integration   $\xi_{out},x_{out}\to\infty$.
The expression
\eqref{energ_disk_relative} gives the relative amount of the total energy emitted at the given frequency,
but not the radial distribution of the energy. That is, it is assumed that a distant observer measures this energy at the given frequency.

\begin{figure}[t]
\begin{minipage}[t]{.5\linewidth}
  \begin{center}
  \includegraphics[width=9.5cm]{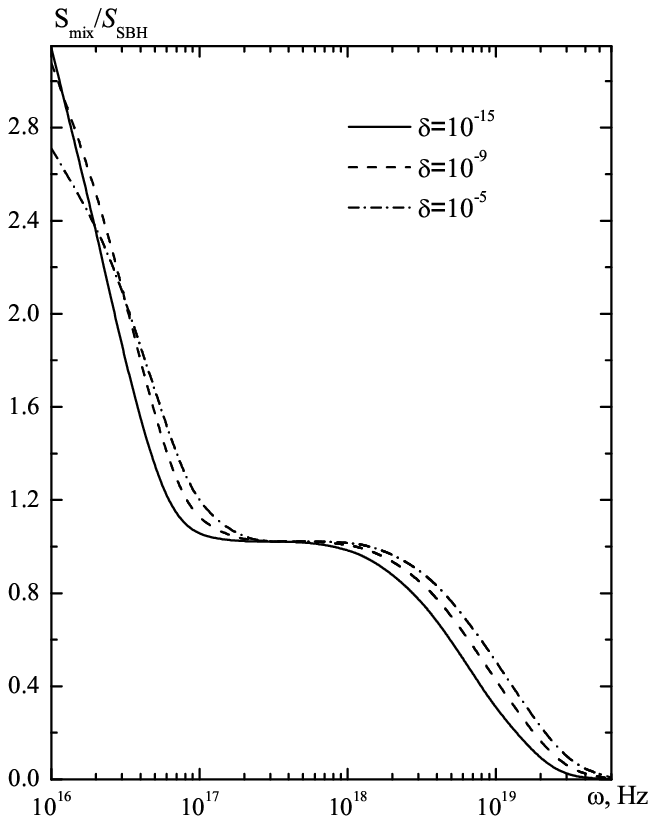}
  \end{center}
\end{minipage}\hfill
\begin{minipage}[t]{.5\linewidth}
  \begin{center}
  \includegraphics[width=9.5cm]{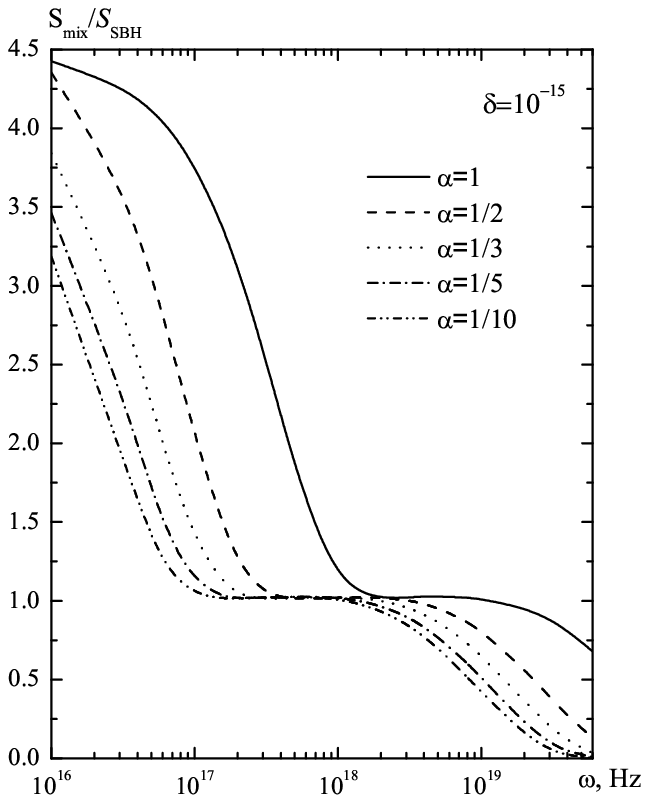}
  \end{center}
\end{minipage}\hfill
\vspace{-.5cm}
  \caption{The total energy emitted from the disk in terms of $S_{\text{SBH}}$ for a mass accretion rate
  $\dot{M}_0=\left(10^{-9}M_{\odot}/\text{yr}\right)\left(M/M_{\odot}\right)$.
The graphs for the system with the polytropic fluid (left panel) and for the system with the Fermi gas (right panel) are shown.
  }
 \label{fig_energ_rad_disk}
\end{figure}

Note that since the mixed systems have a material surface,
then as accreting matter falls onto such a surface, it will
emit a luminosity of the same order as the one emitted by the disk~\cite{Novikov:1973}.
If the total luminosity approaches the ``Eddington limit'',
$L_{\text{Edd}}\sim \left(10^{38} \text{erg/sec}\right) \left(M/M_\odot\right)$,
then radiation pressure will destroy the disk, and the general picture of the
accretion will differ from the one implied by the
standard thin disk model by Shakura \& Sunyaev considered here \cite{Shakura:1972te}.
The latter assumes that the accretion rate is very sub-Eddington.
Therefore we here consider the ``subcritical case'' when  the total luminosity
is much less then $L_{\text{Edd}}$. For this case the accretion rate
$$
\dot{M}_0 \ll \dot{M}_{\text{Edd}}\sim \left(10^{-8} M_{\odot}/\text{yr}\right) \left(M/M_\odot\right).
$$

The results of the calculations from formula \eqref{energ_disk_relative} for the X-ray band are shown in Fig.~\ref{fig_energ_rad_disk}.
For both types of mixed systems, one can see marked differences in the emission spectrum, as compared with SBHs.
In particular, for the system with the polytropic fluid, as $B\to B_{\text{crit}}$ ($\delta\to 0$)
the energy radiated from the system is strongly suppressed at frequencies
 $\omega \gtrsim 10^{18} \text{Hz}$ but demonstrates  a considerable growth at $\omega \lesssim 10^{17} \text{Hz}$.
On the other hand, in the intermediate frequency range $10^{17} \text{Hz} \lesssim \omega \lesssim 10^{18} \text{Hz}$
the curves are flat and lie near unity,
which corresponds to approximate comparability of the radiated energy of
the mixed systems and SBHs.
In turn, the Fermi systems demonstrate a similar behavior:
at the intermediate frequencies, the curves are flat;
in the long-wavelength region of the spectrum,
there is a significant growth of the amount of the radiated energy;
in the short-wavelength region,
the energy of the mixed systems becomes strongly suppressed again.

\section{Discussion and conclusion}
\label{conclusion}

Black hole mimickers are quite popular objects of study. Possessing  strong gravitational fields,
large masses and high-redshift surfaces, for a distant observer, they can look very similar to ordinary black holes.
In addition, solutions describing such systems are regular over all of spacetime, and this is one of their main attractive features.
However, such objects may also possess noticeable differences as compared with BHs. In particular,
since their external spacetime geometry differs from that of BHs, the motion of test particles in gravitational fields
of BHs and BHMs will in general be different.
This manifests itself, for example, when one considers the process of accretion of matter onto such objects.
Then, depending on the particular type of BHMs, both the structure of accretion disks
and their radiant emittance (spectrum) will change.

Thus, if one intends to carry out a more or less complete mimicking of potentially observable characteristics of BHs,
one needs to obtain not just  massive objects possessing high-redshift surfaces, but also to consider other (e.g., astrophysical)
aspects shared by systems of this kind. Only when a proposed object demonstrates a behavior similar to that of a BH with respect
to a certain set of criteria, then it can be regarded as a BHM candidate.

Consistent with this,
in the present paper we have considered  mixed systems consisting of a wormhole (supported by a massless ghost scalar field)  threaded by ordinary
matter (described here in the form of an isotropic perfect fluid).
To model the latter, we have employed two types of EOSs: the relativistic polytropic EOS,
Eq.~\eqref{eqs_NS_WH}, and the EOS for the ideal degenerate Fermi gas,
Eqs.~\eqref{eos_enrg_dens} and \eqref{eos_p}. The distinguishing feature of the mixed systems considered here
is that for the case
when the energy density of the fluid at the center
	   is of comparable magnitude to the one of the scalar field
(i.e., only when the parameters $B$ and $B_1$ are large enough),
the throat of the wormhole can be shifted away from the center of the system. Then the center corresponds to an equator
   surrounded by a double throat. As the amount of ordinary matter increases,
the throats shift further away from the center and can ultimately emerge from the surface of the fluid
(see Fig.~\ref{fig_sigma_nu_exter} and our previous work~\cite{Dzhunushaliev:2014mza}). In this case there are some critical values of
the parameters $B$ and $B_1$ at which physically reasonable solutions no longer exist,
since then the circumferential radius at the center $\Sigma_c \to \infty$.

An interesting feature of such mixed systems is that, as the numerical  calculations indicate,
even as $\Sigma_c \to \infty$ their total masses remain finite. In turn, the value of the redshift function $g_{tt}=e^{\nu_c}$ at the center
goes to zero, which ensures the mentioned finiteness of the mass [in this connection, see the definition of the total mass
via the Komar integral \eqref{mass_Komar_poly} and \eqref{mass_Komar_Fermi}].
Considering systems of this kind, we have shown that
for both types of ordinary matter involved, there exist static, regular, spherically symmetric, asymptotically flat
solutions describing configurations with the ordinary matter concentrated
in the central region.
By choosing
the density of matter at the center and characteristic sizes of the systems in such a way that the parameters
	 $B$ and $B_1$ tend to their critical values, we have obtained
massive objects possessing high-redshift surfaces. For such configurations,
we have found the typical distributions of the energy density (Fig.~\ref{fig_energ_mix})
and of the external gravitational field (Fig.~\ref{fig_sigma_nu_exter}) along the radius.
Also, we have calculated their total masses and the sizes of the fluid surfaces (Figs.~\ref{fig_mass_nu_poly} and \ref{fig_mass_nu_fermi}).
Because of the presence of the latter, such objects, in contrast to BHs, do have a material high-redshift surface
that should result in, for example, additional changes in the emission spectra associated with the process of accretion of matter onto
such a surface (we did not consider such a process here).


In this respect, such objects can be regarded as BHMs. On the other hand, in order to check
their ability to mimic BHs with respect to astrophysically  observable manifestations, we have considered
the process of accretion of test particles onto both types of systems.
For this purpose, we have used the well-known thin accretion disk model of Ref.~\cite{Page:1974he}.
We have shown that for the mixed systems considered here:
\begin{itemize}
\item
The maximum values of the flux of radiant energy flowing out of the disk
are always lower,  reaching $\sim 90\%$ of the maximum for Schwarzschild black holes in all cases.

\item
The fluxes for the mixed systems both with the polytropic and Fermi fluids are very close to those for
the cold black hole.

\item
In general, the minimum radii for a stable orbit
(in units of the gravitational radius) exceed the corresponding value of a SBH.
\item
The efficiency of the conversion of the accreted mass into radiation is about the same as for BHs.
\item
The X-ray blackbody spectrum differs substantially from that of SBHs, see Fig.~\ref{fig_energ_rad_disk}.
\end{itemize}

Thus it is seen that the mixed systems considered here, being similar to SBHs with respect to masses and the presence
of a high-redshift surface, possess specific features manifesting themselves both in the radial distribution
of the radiation flux $F(r)$ and in the emission spectrum of the accretion disk.
By choosing the parameters of the mixed systems, we could not achieve a more or less acceptable coincidence
between the characteristics of their spectra and those of SBHs. In this aspect, apparently, one cannot refer to such systems as
BHMs, as distinct from, for example, the black-hole-like systems of Ref.~\cite{Guzman:2009zz} for which,
by making an appropriate choice of the parameters of boson stars, one can get emission spectra similar to those of SBHs.
On the other hand, when considering systems containing only wormholes~\cite{Harko:2008vy}, the fluxes and spectra  also differ
considerably from those of SBHs. But, as compared with ordinary SBHs, the radiated energy is greater in the case of
pure wormhole systems and smaller for our mixed systems.
It provides an opportunity to
distinguish the external geometry  of such mixed systems both from Schwarzschild and pure wormholes geometries
in astrophysical observations of emission spectra from accretion disks.

In conclusion, we would like to briefly address the question of stability of the mixed systems.
Recent investigations
of Refs.~\cite{Dzhunushaliev:2013lna,Dzhunushaliev:2014mza,Aringazin:2014rva}  revealed that
the mixed systems of the type considered here are unstable.
Obviously,
this instability arises because of the presence of a wormhole
based on a phantom field~\cite{Gonzalez:2008wd,Bronnikov:2011if,Bronnikov:2012ch}.
Possible ways to avoid such an instability could be to consider mixed systems with initially stable
wormholes. These could be (i) wormholes within the framework of modified theories of gravity~\cite{Kanti:2011jz,Kanti:2011yv};
(ii) rotating wormholes, when rapid rotation might favour stabilization of the system~\cite{Dzhunushaliev:2013jja};
(iii) special types of general-relativistic  wormholes \cite{Bronnikov:2013coa}.
In any case, the consideration of mixed systems
with the aforementioned wormholes requires special studies
both with regard to stability
and with regard to the very existence of the required solutions as a whole.

\section*{Acknowledgements}
We gratefully acknowledge support provided by the Volkswagen Foundation.
This work was further supported
by Grant No.~316 in fundamental research in natural sciences
by the Ministry of Education and Science of Kazakhstan,
by the DFG Research Training Group 1620 ``Models of Gravity'',
and by FP7, Marie Curie Actions, People,
International Research Staff Exchange Scheme (IRSES-606096).
V.F. also would like to thank the
Carl von Ossietzky University of Oldenburg for hospitality
while this work was carried out.

\end{document}